\documentclass[sigconf]{acmart}
\usepackage{listings}
\usepackage{xcolor}
\usepackage{graphicx}
\usepackage{amsmath}
\usepackage{listings}
\usepackage{float}
\usepackage{floatflt}
\usepackage{comment}

\renewcommand\footnotetextcopyrightpermission[1]{} 
\begin{document}
\title{Reinforcement Learning Framework for Quantitative Trading}

\author{Alhassan S. Yasin}
\email{ayasin1@jhu.edu}
\orcid{0009-0001-8033-9850}
\affiliation{%
  \institution{Johns Hopkins University}
  \city{Baltimore}
  \state{Maryland}
  \country{USA}
}

\author{Prabdeep S. Gill}
\email{pgill6@jhu.edu}
\orcid{0009-0002-1685-0707}
\affiliation{%
  \institution{Johns Hopkins University}
  \city{Baltimore}
  \state{Maryland}
  \country{USA}
  }

\begin{abstract}

The inherent volatility and dynamic fluctuations within the financial stock market underscore the necessity for investors to employ a comprehensive and reliable approach that integrates risk management strategies, market trends, and the movement trends of individual securities. By evaluating specific data, investors can make more informed decisions. However, the current body of literature lacks substantial evidence supporting the practical efficacy of reinforcement learning (RL) agents, as many models have only demonstrated success in back testing using historical data. This highlights the urgent need for a more advanced methodology capable of addressing these challenges. There is a significant disconnect in the effective utilization of financial indicators to better understand the potential market trends of individual securities. The disclosure of successful trading strategies is often restricted within financial markets, resulting in a scarcity of widely documented and published strategies leveraging RL. Furthermore, current research frequently overlooks the identification of financial indicators correlated with various market trends and their potential advantages.

This research endeavors to address these complexities by enhancing the ability of RL agents to effectively differentiate between positive and negative buy/sell actions using financial indicators. While we do not address all concerns, this paper provides deeper insights and commentary on the utilization of technical indicators and their benefits within reinforcement learning. This work establishes a foundational framework for further exploration and investigation of more complex scenarios.

\end{abstract}


\maketitle
\fancyfoot{}
\fancyhead{}
\thispagestyle{empty}

\section{Introduction}

The primary goal for investors in financial markets has always been the same, minimize trading risk and maximize profits, in the form of returns. Achieving such an objective involves using a systematic approach to predict the price of a security and the type of trends that are to follow. This is a complex and dynamic task, which makes it hard to understand which factors to include and which ones to exclude when making a decision. The research conducted in this sub-field has been primarily around designing automated trading systems which take over the responsibility of fundamental investor while also providing greater liquidity into the financial markets. Both supervised and unsupervised learning techniques have been used in the past, however, they have provided rather unfavorable results. Reinforcement Learning (RL) has gained attention in the last several years in this particular application, potentially offering novel solutions that allow for a more accurate price forecast, thereby allowing the trading agent to successfully interact with its environment to make optimal decisions. Additionally, the dynamic nature of financial data aligns well with the Markov Decision Process (MDP), which serves as the basis in addressing reinforcement learning challenges. The Markov Decision Process helps to evaluate which actions the agent should take considering the observation state and the environment that the agent is interacting with. The Markov Decision Process states that the future state will depend on the current state and action, it assumes all necessary information is captured in the current state. However, this can be problematic in the sense that time-series data is never static. That is to say the Markov Decision Process does not capture all relevant information from a historical perspective. 

In particular, ordinary stock market data consists of open, close, high, low, and volume per security, which is commonly presented in a sequential format over various time intervals. Given the nature of time-series data and its sensitivity to subtle changes, the introduction of technical indicators was proposed to help with creating more robust trading strategies. Furthermore, this helps to identify what characteristics and patterns bullish and bearish scenarios may entail. Today, the challenge remains in effectively utilizing these indicators to reliably forecast future price movements. The presence of false signals adds another layer of complexity into the mix. To address this, advanced techniques such as dimensionality reduction, feature selection, and extraction have been used through Convolutional  Neural Networks (CNN), Recurrent Neural Networks (RNN), Long Short Term Memory (LSTM), and more. 

In the experiments section, the models utilized 20 technical indicators as inputs. A more detailed explanation of many of these indicators can be found in the Technical Analysis book by John J. Murphy \cite{murphy1999technical}.  

\section{Problem Statement and Motivation}
 Reinforcement learning can be leveraged by individuals to develop more robust trading strategies. The competition that exists today makes it hard for individuals to compete due to the limited computational resources. A quantitative hedge fund on the other hand has a wide array of resources that make it easier for them to conduct research, develop, and refine strategies. The core motivation of this paper is primarily due to the lack of knowledge given the limited number of publications in this space. Quantitative hedge funds and even individual investors are hesitant to reveal any trading strategies to the general public as this will eliminate their competitive edge. 

 Our work showcases the considerations one needs to take into account when using reinforcement learning with indicators. The findings in this paper create a foundation examining key components such as data pre-processing, back testing (using Backtesting.py), reward function, normalization methods, and much more. This work is not intended to help individuals with financial gains. A thorough exploration of the limitations are discussed and subject matter experts can further develop this work.

\section{Background Overview of the Work}

\subsection{Background: RL in Quantitative Trading}
Reinforcement Learning (RL) has shown significant promise in the realm of quantitative trading, outperforming other machine learning methodologies in many cases. In this context, RL agents are designed to interact with the financial market environment, receiving feedback in the form of rewards or penalties based on their trading actions. The primary objective of these agents is to maximize their cumulative rewards over time, enabling them to make more accurate and optimal trading decisions. 

A key advantage of RL in quantitative trading is its ability to process and analyze vast amounts of financial data, which can lead to more informed trading strategies.~\cite{vyetrenko2019risk} However, an overabundance of data can sometimes overwhelm the RL agents, leading to less effective decision-making. This necessitates the development of advanced methodologies to filter and prioritize relevant information. Moreover, RL's effectiveness is often measured by its ability to predict future market movements and generate profitable trading strategies. The introduction of a discount factor, $\gamma$ (Figure 1), helps account for the uncertainty of future rewards, balancing immediate gains with long-term potential.


\begin{figure}[H]
\centerline{\includegraphics[width=\columnwidth]{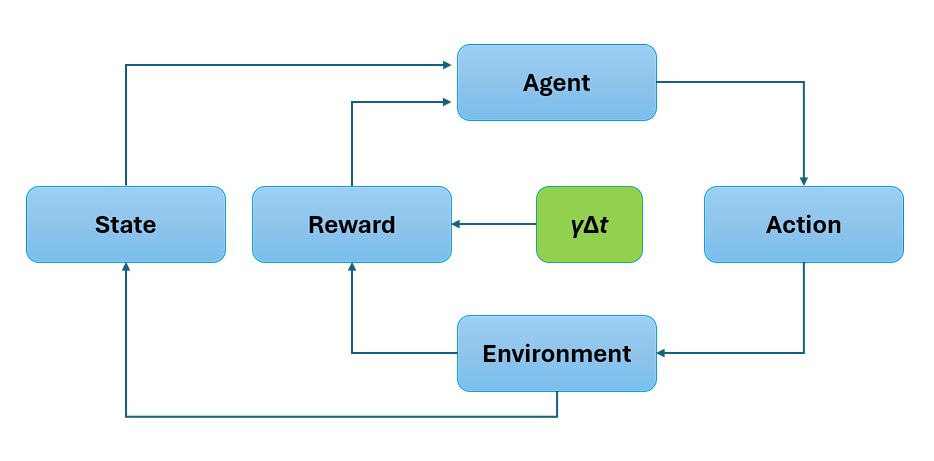}}
\caption{Reinforcement Learning Process}
\label{fig:example}
\end{figure}

\subsection{Background: Literature on Indicators}

A financial indicator in the context of a reinforcement learning agent is a measure that will help provide insights into the future forecast of a security with respect to price, ultimately helping better understanding trends, changes, and rapid fluctuations in the market. Current research does not provide a structured methodology or approach into indicator selection, thus these indicators are selected without conducting an in-depth examination of the pros and cons. This limits the performance of the agent, but also fails to provided the agent with an observation space that is ideal to optimize decision making. Therefore, it is crucial to analyze the impacts of various indicators on separate trading strategies. 

The universal issue with indicators stems from the contrast between short and long time frames. Short-term indicators react quickly to price changes (prone to a larger degree of volatility), while long-term ones smooth out the fluctuations. As the long-term time periods smooth out, it begins to get complicated to identify trends and shifts with precision. Figure 2 showcases this by examining 10 different SMA trends over various time periods. The 10 different SMA's represent the following time periods: (1). 5-days, (2). 10-days, (3). 15-days, (4). 20-days, (5). 25-days, (6). 30-days, (7). 35-days, (8). 40-days, (9). 45-days, and (10). 100-days.

\begin{figure}
    \centering
    \includegraphics[width=1\linewidth]{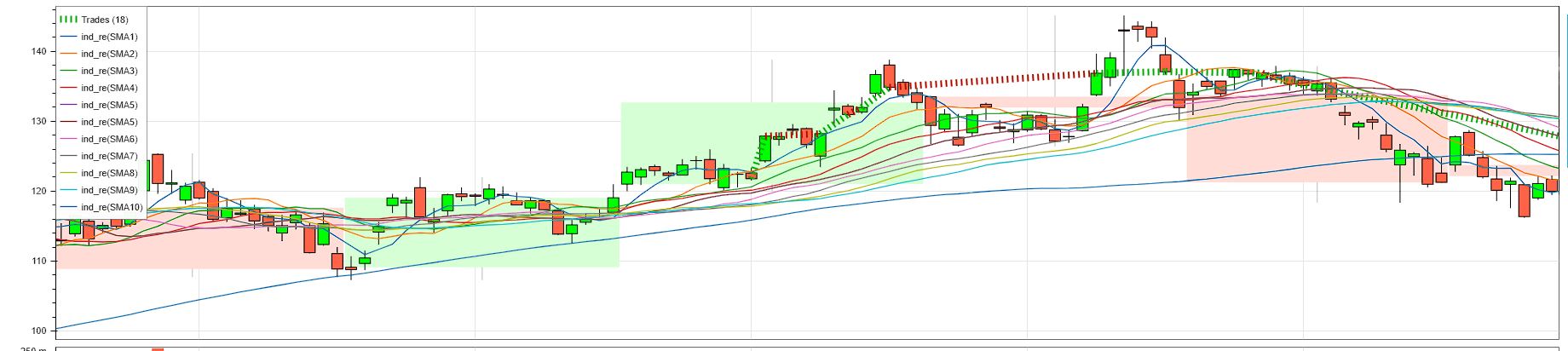}
    \caption{SMA over 10 different time periods}
    \label{fig:enter-label}
\end{figure}

In Figure 3, we zoom in on SMA1, which represents the moving average over 5 trading days and SMA10 representing the moving average over 100 trading days.

\begin{figure}[H]
        \centering
        \includegraphics[width=1\linewidth]{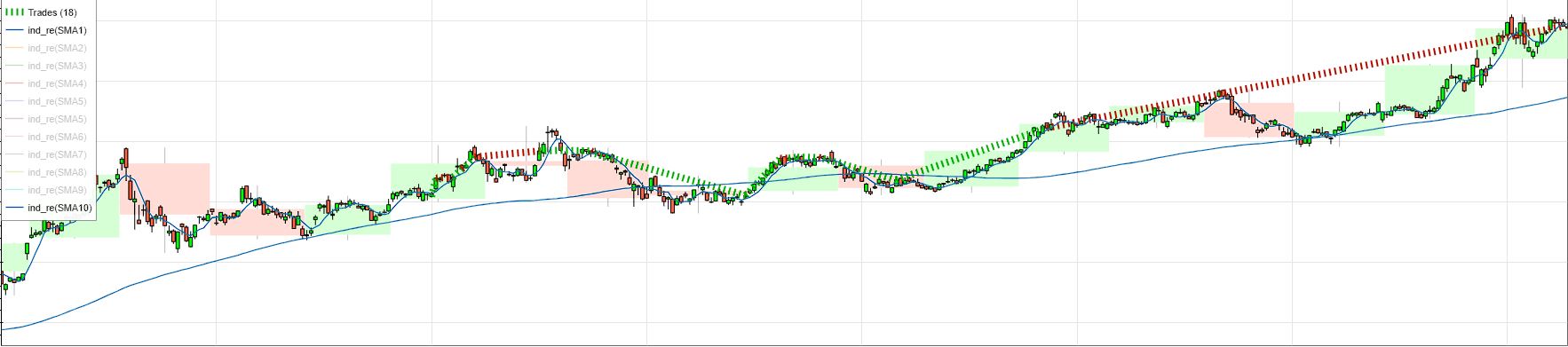}
        \caption{SMA 5-days vs 100-days}
        \label{fig:enter-label}
\end{figure}

Ultimately, it comes down to the type of RL trader and its desired frequency of trading. On an important note, the greater number of trades executed will results in higher commission fees when tying this concept back to a dynamically shifting market. A shorter-term trader is more focused on exploiting volatility for gaining profits, while the longer-term trader utilizes longer time frames in an asset, anticipating a favorable uptrend or increase in value. Figure 3 showcases two separate moving averages, a short-term and a long-term. The short-term moving average follows market fluctuations closely as it is taking into consideration the last 5 trading days. The long-term moving average, shown in dark blue, smooths out and captures a long-term perspective taking into account the last 100 days without consideration to short-term movements.

Careful consideration is required to ensure profitability can be obtained during the trading period as it is rare to see two indicators correctly identify an asset trend. Therefore, it is necessary to leverage and combine different signals to better address the shortcomings associated with the use of a single indicator.

\subsection{Discrete and Continuous Action Spaces }

An overview of discrete and continuous action spaces are important to better understand the problem statement.\cite{vyetrenko2019risk} A discrete or continuous action space can be used in this specific application depending on the desired objective. A discrete action space can be defined as a buy or sell action from the RL agent, similar to what this work has explored. On the contrary, a continuous action space can be applied such that various actions take place defined by a range of 0 to 1. For example, a zero could indicate a sell action and a one could indicate a buy action. Something within the range could indicate the confidence level of a buy or sell. Moreover, within a continuous action space, the range (0 to 1) could be utilized to indicate how much of the total assets in the portfolio should be allocated to a specific security. There are many ways to view and address this exact problem. 

Currently, in reinforcement learning-based stock trading models, the action space is defined as a continuous range from 0 to 1. This is the case in models that utilize Deep Q-Networks, or Policy Gradient methods like Proximal Policy Optimization (PPO) or Trust Region Policy Optimization (TRPO) as their performance has proven to be enhanced utilizing a continuous action space in this respect. \cite{schulman2015trust} Furthermore, this problem can be viewed from the portfolio management perspective as well. \cite{guan2021explainable} The continuous action space can be represented such that a value of 0 will indicate that the RL agent does not buy that stock (or sell if shares are owned), while a value of 1 can indicate that the RL agent allocates say some percent of the overall portfolio. The application of utilizing a continuous action space is versatile as it can be used to address a  portfolio management problem and can also be used to adapt a singular buy/sell model as well.   

\section{Literature Review}

A study published in 2016 used a deep reinforcement learning approach without relying on traditional technical indicators. \cite{deng2016deep} By restructuring a Recurrent Deep Neural Network (RDNN), the study enabled simultaneous environment perception and recurrent decision-making in real-time trading. Deep Learning was utilized for feature extraction and other fuzzy learning principles helped to handle input data.  While the solution did achieve profitability and outperformed conventional strategies in various market conditions and in certain time periods, the study’s evaluation was limited to a small set of contracts. The study was not able to provide evidence of the strategy performance in live trading. 

Another study integrated deep reinforcement learning along with sentiment analysis to demonstrate its efficacy in learning stock trading strategies. \cite{azhikodan2019stock} This two-prong approach helps to minimize the uncertainty associated with pure quantitative trading strategies using only algorithms. The approach outlines employing DDPG for the reinforcement learning agent and an RCNN for sentiment analysis to help with enhancing the model’s predictive capabilities. The experiments conducted do not demonstrate a significant increase in profits when comparing against the market average, but they do analyze the agent’s ability to learn trading strategies. There are also some limitations in the reward function as it does not take risk factors into account. The reward function is binary such that the agent receives a reward of 1 if the action taken is profitable and a 0 if the action is not. While this may be effective in the short-term there is no methodology that can prove it’s outcome to be effective in long-term trading. 

Another study introduced a novel reinforcement learning framework for financial portfolio management, one that doesn’t rely on financial models. \cite{jiang2017deep} The framework is made up of the Ensemble of Identical Evaluators (EIIE) topology, a Portfolio-Vector Memory (PVM), an Online Stochastic Batch Learning (OSBL) scheme, and an explicit reward function designed to maximize returns and manage risks effectively. Implemented with Convolutional Neural Network (CNN), a basic Recurrent Neural Network (RNN), and Long Short-Term Memory (LSTM). The framework is tested with three back-test experiments in a 30-minute trading window within the cryptocurrency market. While the results showcased impressive returns, the study merely focused on the cryptocurrency market and not traditional financial markets. 

This approach evaluates nine different environmental variables, which includes open, close, high, low, previous day’s closing price, trading volume, turnover, percentage change, and absolute change in stock price. \cite{xu2023deep} The paper introduces three main states for the trading agent: Buy, Sell, or Hold. It highlights the effectiveness of the double average trading strategy in assisting agents to take actions that will be associated with the highest return/reward. Both short-term and long-term moving averages are used to determine optimal buying and selling points in the market. Furthermore, the vast majority of research has showcased models that operate best with a long-term returns perspective, the short-term is far more difficult to deliver optimal results consecutively. \cite{li2022stock}


\subsection{Considerations}

 Normalizing and scaling different environmental variables create potential challenges such that each variable would need to contribute equally to the agent’s decision-making process. Inconsistent scaling can lead to an inaccurate model in the long-term. Changing any of these indicators can also potentially influence the rewards received by the agent, requiring careful consideration and adjustment of the reward function itself. \cite{li2022stock} This is another example of a paper that conducted rigorous back testing to ensure optimal results, but the question lies in determining what the hold state entails. 

When we think of reinforcement learning and discrete action spaces in particular, it should be noted that RL agent doesn't exactly understand the difference between a hold state and a do-nothing state. The agent does not understand what hold means unless that is specifically provided as an input to the agent. However, if the agent is not provided this as an input, then this is likely a do-nothing state, as opposed to a hold state. Do nothing is problematic in the sense that the agent does not understand whether to be in a buy or sell action, or maybe even requires more data before being able to take make an optimal decision. It becomes even more complex when considering if the agent is currently holding any assets or not, as that is essentially a do-nothing state when no assets are held at any given timestep. This is different than the agent already having assets and deciding to hold those assets. It was unclear whether the hold state presented any value as the agent must effectively examine the trade-off between gains and risks for a dynamic and ever-changing dataset.

\section{Methodology}
Our work explores potential methods such that reinforcement learning agents can continue to learn and adapt to the environment as the space changes. We investigate how a reinforcement learning agent can utilize financial indicators in specific market conditions and trends to enhance overall trading accuracy. By understanding the correlation between various indicators, we can selectively identify which indicators provide new information to the RL agent. 

Our data pre-processing will consist of exploring multiple ways to normalize reinforcement learning inputs, which will help scale the state space or inputs features of reinforcement learning agents between a specified range.

Our approach considered various reward functions which included additive rewards, terminal rewards, immediate rewards, and final rewards. By experimenting and testing different reward types, the study identified which reward scheme would help the agent make the most optimal trading choices and prioritize recent performance through the last time step. Ultimately, the reward will influence the agent's behavior and frequency of trading. Our goal is to examine and provide insights into both the short-term and long-term trading perspective.  

By narrowing our agent’s focus to two actions: buy or sell, we are potentially able to make it more effective. Next, the study selected three seperate algorithms that would best align with a discrete action space. These included Deep Q-Network (DQN), Proximal Policy Optimization (PPO), and Actor-Critic (A2C).

\subsection{State Space}
The state space in the trading environment consists of historical price data and technical indicators. The dimensionality is determined by the length of the historical data window and the number of indicators. The historical data window is pulled using a Yahoo Finance API, which helped to specify a period of interest.

\subsection{Action Space}
The action space was defined using gym-anytrading, a library that extends the OpenAI Gym framework to provide environments for simulating and testing algorithmic RL trading strategies. By default, this library makes the action space discrete such that the actions consist of either buy or sell actions. The reinforcement learning agent will always be in the market and taking one of those actions at all times. The agent does not occupy a hold or do-nothing state.  

\section{Reward Function}

Extending on the definition of the action space above, the reward function calculates the reward based on the action taken by the agent. Specifically, the reward is positive if the action results in a trade that generates profit and negative if the action results in a loss. The study proposed three potential reward functions to train the RL agent to trade, which included immediate, terminal, and final rewards. 

The first reward function (immediate), evaluates the log price difference of the current time step by the previous time step. It checks if the next action is initiated based on the current position and then will calculate the price difference between the current price and the last trade price. If the current position is a long position, the price difference is added to the step reward. 

The equation can be written as: 

\[
\log\left(\frac{p_i}{p_{i-1}}\right)
\]

The second reward function initializes the reward to be zero. The reward is specifically calculated when there is a change in the action (long/short). If the current action is the same as the previous action, the reward will remain zero. If the current action is not the same as the previous action, the reward is calculated as the natural log of the ratio of the current price to the initial price.  

In the third reward function, the study proposed to provide zero immediate rewards and only provide a single final reward based on the initial equity value versus the final equity value. 

However, all of these reward schemes have limitations.  In this work we primarily focused on immediate rewards as this provided the RL agent to behave an a manner that would maximize the financial gain on each trade. 

\subsection{Reward Function Limitations}
Three separate reward functions were discussed in the previous section. The first reward function was simple, taking the difference between the previous price and current price. However, the problem with giving immediate rewards to the RL model is that it tends to optimize for the smallest possible gains on each trade, rather than considering long-term benefits. Providing an immediate reward at every time step makes the model more prone to issues such as over fitting and/or greater variance in learning. 
Similarly, providing the model with only a final reward means the model will struggle with distinguishing a good trade versus a bad trade. This limits the model's learning ability as it is not optimized to perform well in the future. Careful consideration in structuring the reward function is crucial for developing an RL model, whether it's designed for the short-term trading or long-term.

\section{Utilizing Technical Indicators}

Our work examines and assesses 20 different technical indicators (using TA-Lib) as RL inputs and normalizes them utilizing a min/max scalar. The indicators that were used included SMA, OBV, Momentum, Stochastic Oscillator, MACD, CCI, ADX, TRIX, ROC, SAR, TEMA, TRIMA, WMA, DEMA, MFI, CMO, STOCHRSI, UO, BOP, and ATR. The work uses DummyVecEnv to vectorize the custom environment. This vectorized environment helps the RL algorithms with the execution of multiple environments, which can lead to faster convergence and greater efficiency. The DummyVecEnv is a vectorized wrapper for multiple environments, allowing each environment to be called in sequence on the Python process. 

\section{Normalization Methods}
Four separate experiments were conducted to measure the pattern differences between normalizing all indicators with Min-max, Z-Score, Sigmoid, and L2.

\subsection{Min-max}
The Min-max method is a popular and widely used generic normalization method, where the method scales values of a feature between 0 and 1. Simply subtract the minimum value of the feature from each value and then divide by the range. 

\[
\text{Min-max Normalization} = \frac{x - \min(x)}{\max(x) - \min(x)}
\]

\subsection{Z-Score}

\[
\text{Z-Score Normalization} = \frac{x - \mu}{\sigma}
\]

\textbf{Where:} 
    \begin{itemize}
        \item $x$: Represents the original value
        \item $\mu$: Represents the mean of the data 
        \item $\sigma$: Represents the standard deviation of the data
    \end{itemize}

\subsection{Sigmoid}
The sigmoid normalization method involves transforming micro array expression values using a sigmoid function. The function scales the values between 0 and 1 then adjusts the values based on their distance from the mean and the variability of the data by sample standard deviation. 

\[
\text{Normalized Value} = \frac{1}{1 + e^{-(x - \mu)/\sigma}}
\]

\subsection{L2}
The L2 approach will calculate the square root of the squared values of the vector element. 

\[
\text{Normalized Vector} = \frac{{\text{vector}}}{{\sqrt{\sum_{i=1}^{n} (\text{vector}[i])^2}}}
\]

\subsection{Window Normalization}
A separate function was created to help retrieve a window of signal features, normalizes them using a log transformation scaled by 10, and then return an observation. 

\[
s_{ij}' = \log\left( \frac{s_{ij}}{s_{00}} \right) \times 10
\]

\subsection{Summary}
While our study observed various data pre-processing methods, their impact on pattern recognition and indicator relationships appeared to be marginal. However, the notable potential emerged in the context of utilizing PPO, A2C, and DQN algorithms for training three different models. Investigating how these algorithms interpret these relationships and their influence on executing trades is explored in the next section.

\section{Indicators Correlation Matrix}

\begin{figure}[H]
    \centering
    \includegraphics[width=1\linewidth]{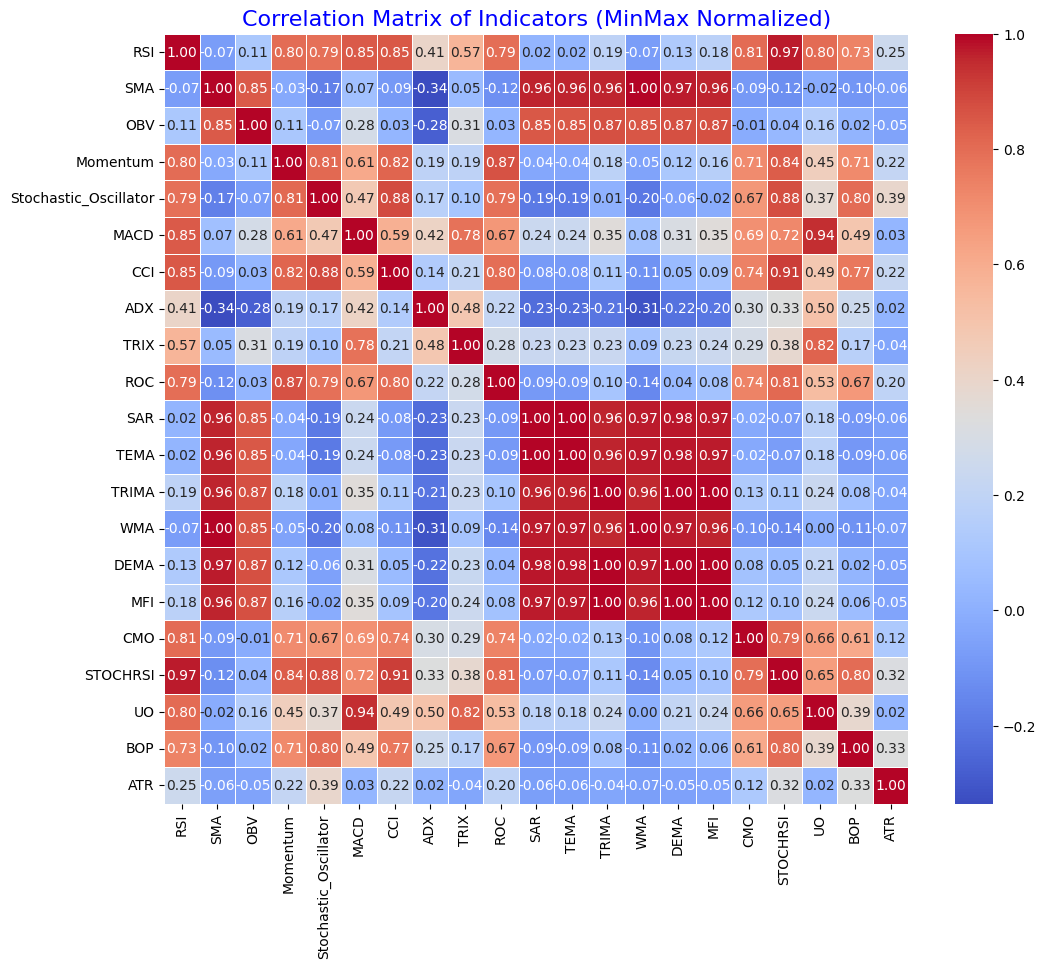}
    \caption{MinMax Normalization}
    \label{fig:min-max_N}
    \vspace*{-10pt}
    \hspace*{-0.4cm}
    
\end{figure}

This correlation matrix in Figure \ref{fig:min-max_N} provides coverage and insights into the relationships (highly correlated or not highly correlated) between the technical indicators in our analysis. Specific patterns include RSI showing strong positive correlation with other indicators including Momentum, Stochastic Oscillator, MACD, CCI, ROC, CMO, STOCHRSI, and UO. This suggests that these indicators move in the same direction as RSI. Other indicators such as ADX, TRIX, SAR, TEMA, TRIMA, WMA, DEMA, MFI, BOP, and ATR show weak correlations with the majority of the other indicators. 

There were three other normalization methods tested on these indicators to gain a better insights into how the correlation relationships would change. Results concluded that all three of these normalization methods provided similar information as the MinMax method. Figure \ref{fig:min-max_N} showcases the level of correlation of indicators measured, all normalized on the same Min-max scale. It should be noted that this may be an ineffective way to measure the correlation relationships among all technical indicators. For example, a momentum indicator may be normalized differently than a volume indicator. 

\section{Training Data}
The initial training dataset was compiled from two years of day data for APPL from 2020-01-01 to 2022-01-01. This data was used to train each of the three models, each uniquely utilizing a separate algorithm. The study zoomed in on Actor-Critic, Proximal Policy Optimization, and the Deep Q-Network as these algorithms have been proven effective in past studies and work well with our discrete action space. Each of the three models were trained on one million time steps.

The agent and the environment engage in a series of interactions through a sequence of time steps. Then at each time step \textit{t},  the agent will receive the state of the environment and a reward that is associated with the previous action. The agent will then take an action. 

\section{Experiments and Results}
\subsection{Actor-Critic}

\begin{figure}[H]
    \centering
    \includegraphics[width=1\linewidth]{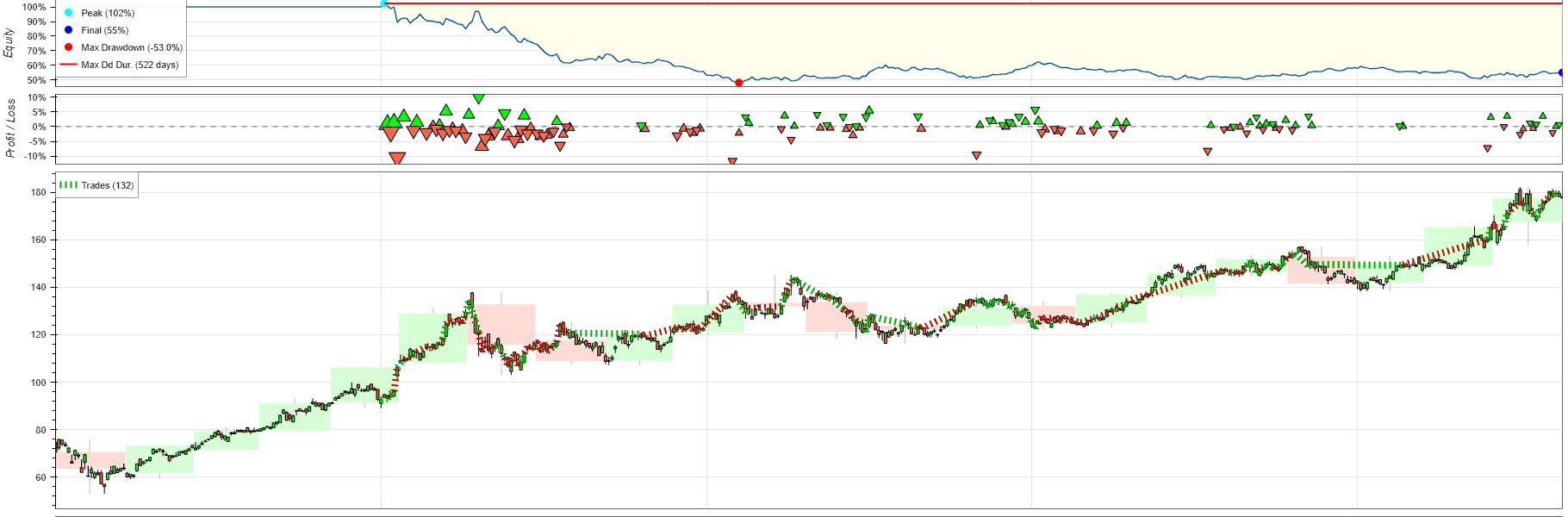}
    \caption{A2C Performance (2-year period), trading APPL, utilizing MinMax normalization with 20 technical indicators as RL inputs, and MlpPolicy}
    \label{a2c}
\end{figure}

As shown in Figure \ref{a2c}, over a two-year period, the A2C approach utilizing the MlpPolicy performed poorly when comparing against the models utilizing DQN and PPO. One of the inherent problems with A2C is that a large sum of data is required for the model to learn effectively. A2C is also prone to suffer from convergence issues, particularly because it is based on gradient descent. This means the parameters are updated following the steepest improvement in overall performance. With respect to time-series data, A2C poses many problems and its performance is able to verify just that.  

\subsection{Proximal Policy Optimization}

\begin{figure}[H]
    \centering
    \includegraphics[width=1\linewidth]{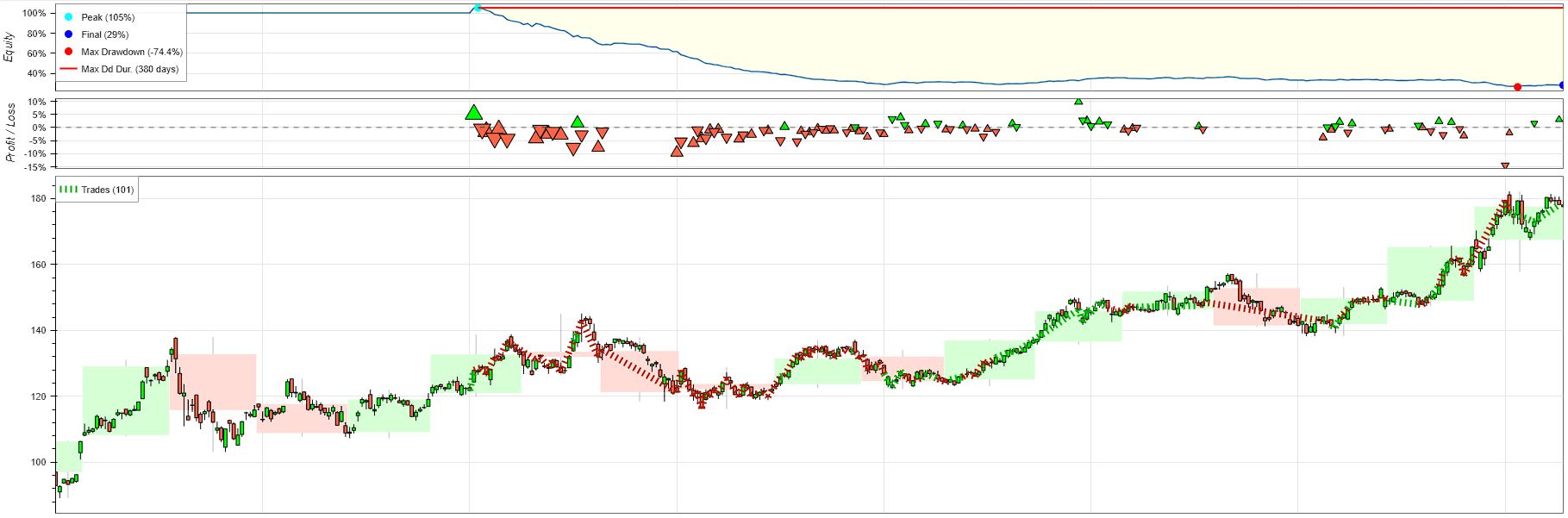}
    \caption{PPO Performance (2-year period), trading APPL, utilizing MinMax normalization with 20 technical indicators as RL inputs}
    \label{fig:PPO}
\end{figure}

The Proximal Policy Optimization (PPO) algorithm \cite{schulman2017proximal} had trouble recognizing when to initiate a buy action or a sell action. There are certain instances where it made reasonable trades, but ultimately failed to clearly distinguish a profitable buy or sell action. As shown in Figure \ref{fig:PPO}, the PPO model made the greatest number of trades within the 2-year period compared to the other algorithms. 

In Figure \ref{fig:PPO}, the PPO model made a total of 101 trades. The PPO algorithm was not effective as the Win Rate was 27 percent across all trades and providing a negative overall return. The annual volatility percentage was at nearly 10 percent and the profit factor produced by the back test was not positive.    

\begin{figure}[H]
    \centering
    \includegraphics[width=1\linewidth]{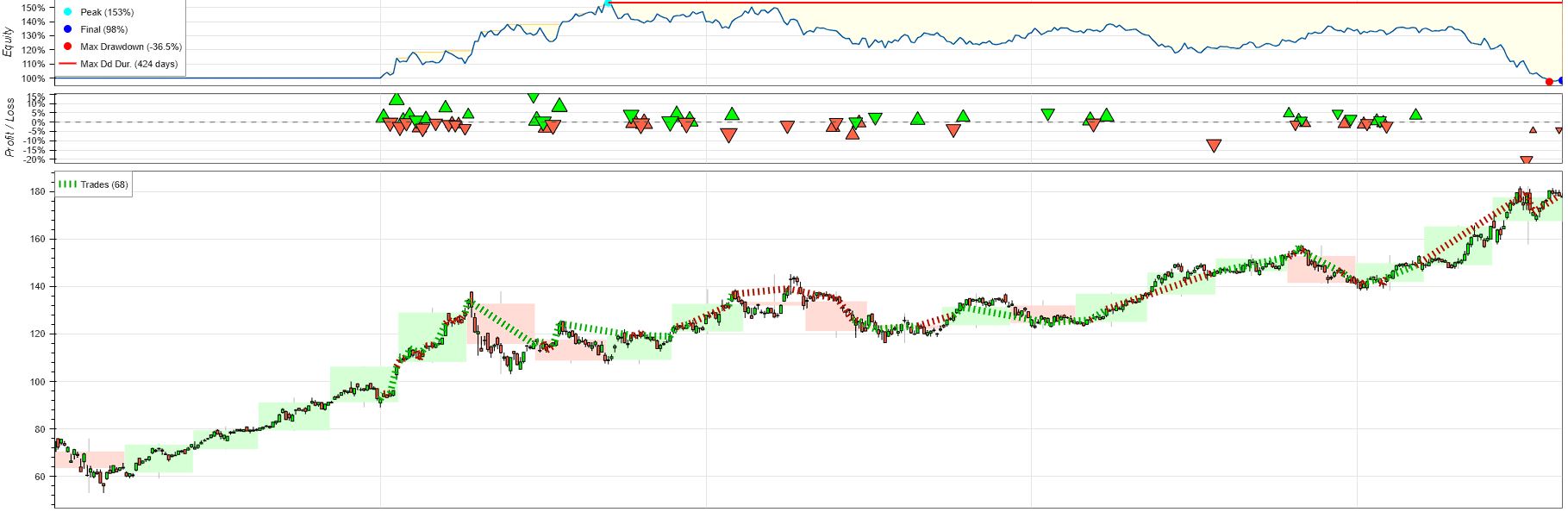}
    \caption{DQN Performance (2-year period), trading APPL, utilizing MinMax normalization with 20 technical indicators as RL inputs, w/o hyper parameters tuning, and MlpPolicy.}
    \label{figure: DQN}
\end{figure}

The Deep Q-Network (DQN) \cite{mnih2013playing} model yielded compelling evidence over a period of several weeks, resulting in a series of profitable buy actions that increased returns. However, its performance was somewhat erratic during certain weeks. Additionally, it frequently executed buy and sell actions in close proximity to each other, indicating that the strategy may lack stability. Given this model gave us the best performance thus far, evaluating it further with hyper parameter tuning was initiated. he DQN model made a number of profitable trades, in the beginning it did well in understanding what constituted a good trade, however, it did struggle in specific windows of time.

Figure \ref{figure: DQN} showcases how the DQN performs over a 2-year period with changes to the hyper parameters. As the learning rate controls the step size during gradient descent, lowering the learning rate to attempt to help stabilize the model.

\subsection{Hyper Parameters}

\begin{table}[htbp]
\centering
\caption{Hyper parameters}
\label{fig: hyperparams1}
\begin{tabular}{|l|l|}

\hline
\textbf{Hyper parameter} & \textbf{Value} \\ \hline
Learning Rate            & 1e-4           \\ \hline
Buffer Size              & 100000         \\ \hline
Batch Size               & 128            \\ \hline
Gamma                    & 0.99           \\ \hline
Target Update Interval   & 1000           \\ \hline
\end{tabular}
\end{table}

Table \ref{fig: hyperparams1} showcases the five different hyper parameters being examined. To better assess the model's ability to explore and learn, the study tested different learning rates to identify any patterns or changes with respect to the performance of the training set.

\subsection{Learning Rate}
Next, the study tested different learning rates, denoted as \( \alpha \). specifically on the DQN model as it was the most stable model during training. Initial tests on training data concluded that utilizing a larger learning rate helped the agent with overall performance compared to a smaller learning rate. Utilizing a larger learning rate helps with exploration, however, this learning rate can be modified to become smaller after the model shows greater evidence of stability.

\subsection{Hyper Parameter Results}

\begin{figure}
        \centering
        \includegraphics[width=1\linewidth]{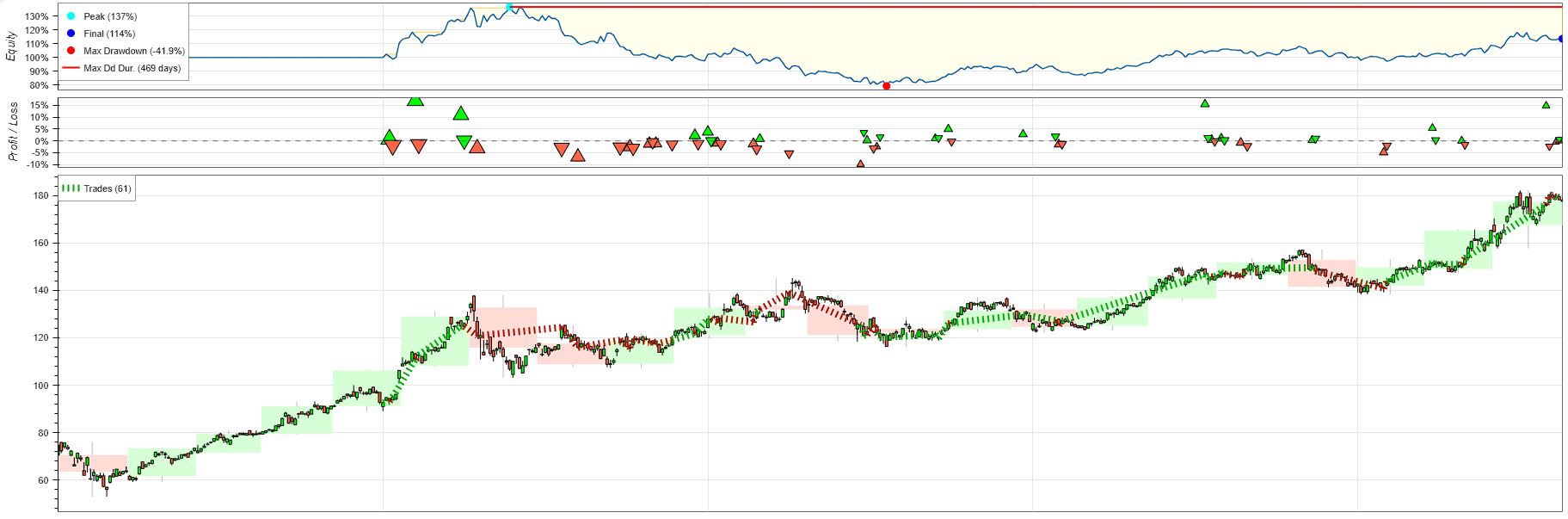}
        \caption {DQN Performance with le-4 Learning Rate}
        \label{fig:DQN w/ Modified Learning Rate}
    \end{figure}

Figure \ref{fig:DQN w/ Modified Learning Rate} showcases utilizing DQN with multiple changes, adding parameters such as learning rate, buffer size, batch size, gamma, and target update interval. For example, the batch size is the number of experiences that are sampled from the buffer and then used to update weights during the iterations. After changing the learning rate to le-4, the model began to better distinguish profitable trades. Table \ref{tab:performance_stats} reflects this as there was a 13.5 percent return over the same 2-year period.  As the table presents many different results, identifying the return percentage, volatility, Sharpe ratio, and win rate are the most important metrics to examine in our evaluation. 

\begin{table}[htbp]
\centering
\caption{Results}
\label{tab:performance_stats}
\begin{tabular}{|l|l|}
\hline
\textbf{Results}          & \textbf{Value}       \\ \hline

Return [\%]                  & 13.558111             \\ 
Return (Ann.) [\%]           & 7.149259             \\ 
Volatility (Ann.) [\%]       & 29.531209             \\ 
Sharpe Ratio                 & 0.242092             \\ 
Sortino Ratio                & 0.392577             \\ 
Calmar Ratio                 & 0.170546             \\ 
Win Rate [\%]                & 45.901639            \\ 

                                                     \hline
\end{tabular}
\end{table}

\subsection{Results continued}

In the next experiment, the learning rate was modified from le-4 to le-2, while all other parameters were left unchanged as shown in Table 3. The difference between these two learning rates is that le-2 promotes quicker convergence as the learning rate is higher and the network will adjust weights much faster. Conversely, a smaller learning rate, such as le-4 will provide more stable updates to the weights. Given the updates are smaller, the overall training time does increase when utilizing a smaller learning rate. Better understanding the trade-offs between exploration and exploitation is what the study explored in this portion of the paper.  
\begin{table}[htbp]
\centering
\caption{Hyper parameters}
\begin{tabular}{|l|l|}

\hline
\textbf{Hyper parameter} & \textbf{Value} \\ \hline
Learning Rate            & 1e-2           \\ \hline
Buffer Size              & 100000         \\ \hline
Batch Size               & 128            \\ \hline
Gamma                    & 0.99           \\ \hline
Target Update Interval   & 1000           \\ \hline
\end{tabular}
\end{table}

\begin{figure} [H]
    \centering
    \includegraphics[width=1\linewidth]{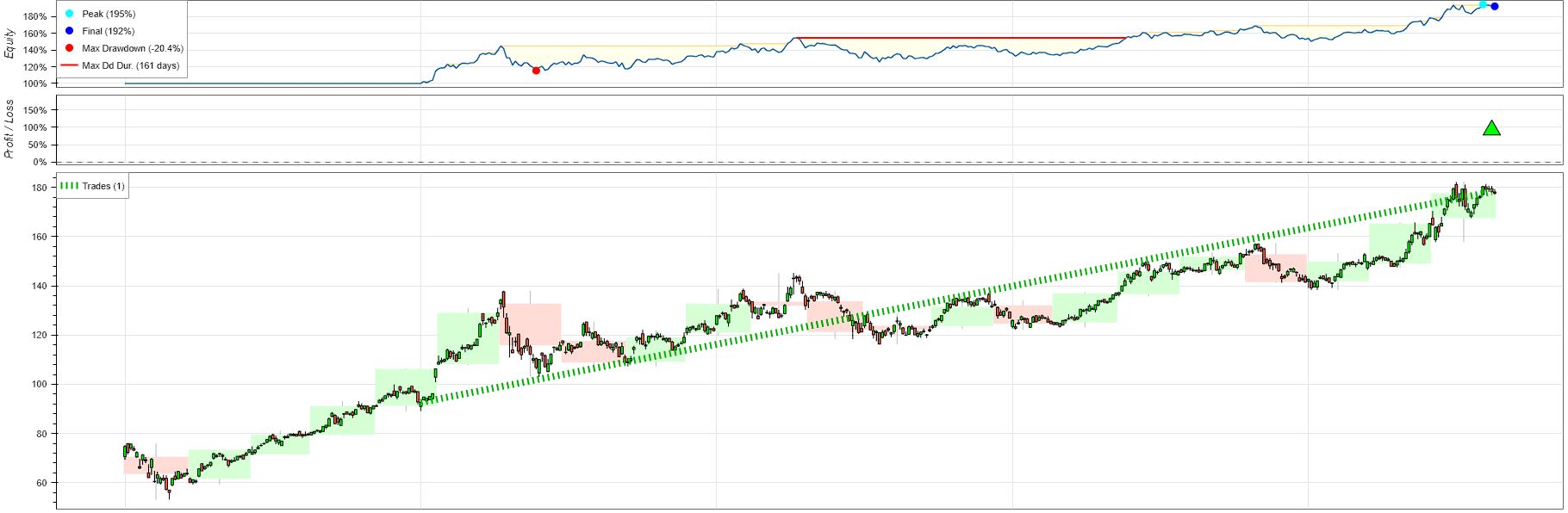}
    \caption{DQN Performance with le-2 Learning Rate}
    \label{fig:enter-label}
\end{figure}

Table \ref{tab:performance_stats2} showcases the performance of the DQN model while changing the hyper parameters.The key metrics to pay attention to are the win rate, Sharpe ratio, return, and volatility Specifically, the learning rate was changed during the study from le-4 to le-2. This simple change made a significant difference among return, Sharpe ratio, and the win rate. The Sharpe ratio is a method for measuring risk adjusted return. 

Calculated as follows: 
\[
\text{Sharpe Ratio} = \frac{R_p - R_f}{\sigma_p}
\]
where \(R_p \) is the Expected Portfolio Return, \(R_f\) is the Risk-Free Rate, and \(\sigma_p \) is the Standard Deviation of Portfolio (Risk). 

\begin{table}[htbp]
\centering
\caption{Results}
\label{tab:performance_stats2}
\begin{tabular}{|l|l|}
\hline

\textbf{Results}          & \textbf{Value}       \\ \hline

Return [\%]                  & 92.316259             \\ 
Return (Ann.) [\%]           & 42.642871            \\ 
Volatility (Ann.) [\%]       & 39.851569             \\ 
Sharpe Ratio                 & 1.070042              \\ 
Sortino Ratio                & 2.395635             \\ 
Calmar Ratio                 & 2.094297             \\ 
Win Rate [\%]                & 100.0            \\ 

                                                     \hline
\end{tabular}
\end{table}

Interestingly, the model only traded a single time compared to the 61 trades while utilizing a larger learning rate. However, this brought up concerns around the potential over fitting problem as the model exhibited  substantial changes in the win rate percentage and Sharpe ratio both increasing. 

The results showcase the superior performance of DQN in comparison to the other two algorithms used. Given the large amount of data the model was processing, it was important to investigate hyper parameters and how their impact on the model's stability and performance. Specifically, the buffer size determines the number of experiences that need to be gathered before performing gradient descent on the entire set. The learning rate also influenced the model's ability to learn more effectively. While our work only performed a few experiments with tuning parameters, much of this work can be expanded with additional experiments around hyper parameters alone. 

\subsection{INTERVALS \& STRATEGY DEGRADATION}
The current work here presents several avenues for enhancement and deeper exploration. Zooming in to one particular area is data intervals. Further broadening the scope of testing towards more granular data intervals such as hourly, minutes, or seconds can help capture more intricate details and provide greater insights into the performance of each of the trained RL agent's. Additionally, it is important to conduct a comparative analysis among the different models in varying market conditions. For example, investigating whether DQN outperforms PPO during downtrends and understanding the underlying reasons behind such performances could yield better insights. Furthermore, incorporating a diverse range of policies from Stable-Baselines3 alongside hyper-parameter tuning, exploration/exploitation strategies are crucial steps required to refine these models. Models will need to be continuously refined as the financial markets provide substantial amounts of new data during trading hours. 

Another universal limitation is strategy degradation. While a single strategy may be back tested using historical data, the future performance of a strategy will always be uncertain. Utilizing a more comprehensive approach such as an ensemble approach helps to minimize the risk associated with training the RL agent on a single algorithm, but does not adequately address the problem with strategy degradation. 

\section{strategy development}
When developing RL trading strategies, there are many considerations one will need to make. This work provides some of the key areas to consider when developing RL trading strategies. An individual can immediately replicate this work and begin to explore with different stock tickers. There are many strategies that can be developed and tested using Backtesting.py and setting stop losses, order sizing, and other specific trading features can help simulate a real-world trading environment utilizing RL. One can further expand on using all 200+ technical indicators as inputs and identify the best normalization schemes for those indicators. For example, normalization will be different for Moving Average Convergence Divergence (MACD), a lagging indicator,  as the range is between -1.0 and 1. A Max-Min normalization method scales all the data between 0 and 1, which can become problematic. One should carefully select indicators and normalize them appropriately to limit results from becoming distorted. 

\section{Conclusion}

Financial markets are characterized by their inherent volatility and dynamic fluctuations, making it crucial for investors to adopt comprehensive strategies that incorporate risk management, market trends, and movements of individual securities. Current literature aims for robustness and highlighting the need for advanced methodologies to bridge the gap for reinforcement learning (RL) in quantitative trading. Key contributions of this work for enhancing foundational framework for reinforcement learning in financial trading:
\vspace{-0.09cm}
\begin{itemize}
    \item Emphasis on Technical Indicators: This work underscores the importance of financial indicators in enhancing RL agents' decision-making processes. By leveraging these indicators, investors can better understand market trends and make more informed decisions.
    
    \item Reinforcement Learning as a Promising Approach: The paper advocates for RL as a potent tool in quantitative trading, noting its ability to handle the complexities and dynamic nature of financial data. The alignment of financial data with the Markov Decision Process (MDP) is particularly noteworthy, as it allows the agent to make optimal decisions based on the current state and actions.
    
    \item Challenges and Opportunities: While RL shows promise, this paper highlights the challenges, such as the RL agent's potential to be overwhelmed by excessive data, which can impede its ability to recognize patterns and make favorable decisions. Addressing these challenges is crucial for developing more effective trading strategies.
    
    \item Innovative Methodology: By focusing on the practical application of financial indicators within RL frameworks, this research provides a foundational framework for improving the performance of RL agents. This involves:
    \begin{itemize}
        \item Data pre-processing and normalization methods
        \item Backtesting using tools like Backtesting.py
        \item Analyzing the impacts of various indicators on trading strategies
    \end{itemize}
    
    \item Future Directions: The insights from this work serve as a stepping stone for further exploration and refinement of RL in financial trading. The goal is to bridge the gap between theoretical models and real-world applications, ensuring that RL agents can make accurate and beneficial trading decisions.
\end{itemize}

Throughout the experiments conducted, the work leveraged three different types of algorithms (A2C, PPO, and DQN) to test performance and experiment with tuning the hyper parameters to help increase stability and performance. Limitations such as strategy degradation and training on a limited set of data do exist. Therefore, it is crucial to test on a wide range of data and to recognize when a strategy is deteriorating.

\bibliographystyle{ACM-Reference-Format}
\bibliography{bib}

\newpage
\clearpage

\end{document}